\documentclass[prl,twocolumn,notitlepage,superscriptaddress,longbibliography]{revtex4-1}
\usepackage{amsmath}
\usepackage{amssymb}

\usepackage[unicode=true,colorlinks=true,citecolor=blue,urlcolor=blue]{hyperref}

\usepackage{bm}
\usepackage{epsfig}

\usepackage[normalem]{ulem}
\renewcommand {\Im}{\mathop\mathrm{Im}\nolimits}

\newcommand{\GO}{\gamma_{\rm 1D}}

\renewcommand {\phi}{{\varphi}}
\newcommand {\rmi}{{\rm i}}

\newcommand {\e}{{\rm e}}
\newcommand {\eps}{\varepsilon}

\begin{document}
\title{%
Dimerization of many-body subradiant states in waveguide quantum electrodynamics
}

\author{Alexander V. Poshakinskiy}
\affiliation{Ioffe Institute, St. Petersburg 194021, Russia}

\author{Alexander N. Poddubny}
\email{poddubny@coherent.ioffe.ru}

\affiliation{Ioffe Institute, St. Petersburg 194021, Russia}

\begin{abstract}
We study theoretically  subradiant states in the array of atoms coupled to photons propagating in a one-dimensional waveguide focusing on the strongly interacting many-body regime with large excitation fill factor $f$.
We introduce a generalized many-body   entropy of entanglement based on exact numerical diagonalization followed by a high-order singular  value  decomposition. This approach has allowed us to visualize and understand the structure of a many-body quantum  state. We  reveal the breakdown of fermionized subradiant states with increase of $f$ with emergence of short-ranged dimerized antiferromagnetic correlations at the critical point $f=1/2$ and the complete disappearance of subradiant states at $f>1/2$.
\end{abstract}
\date{\today}

\maketitle
{\it Introduction.}  Dicke model, describing collective radiance of light-coupled dense atomic clouds, is one of the paradigmatic concepts of quantum optics~\cite{Dicke1954,Scully2009}. Recently, it has become possible to test the classical ideas of collective  spontaneous emission for man-made platforms of waveguide quantum electrodynamics (WQED), studying arrays of natural or artificial atoms (superconducting qubits, quantum dots, quantum defects) coupled to photons propagating in a waveguide~\cite{Roy2017,KimbleRMP2018,sheremet2021waveguide}. 

Historically, the research has been mainly focused on the superradiant symmetrically excited Dicke states of atomic arrays~\cite{Feld1976,Yudson1984}. All other states of the model, which appear to be subradiant, attracted attention only recently.  The structure of subradiant states is much more subtle due to  intrinsically high $(C_N^k-1)$-fold degeneracy of their spectrum  (here $N$ is the number of atoms, $k$ is the number of excitations).  Such  large number of subradiant states can make the spontaneous decay dynamics strongly non-exponential~\cite{Chang2019,Masson2020}.  Single-excited ($k=1$) subradiant states are relatively simple, they can be constructed as a superposition of individual atom excitations that is  out of phase with the light wave. However, subradiant states with $k>1$ excitations have become a subject of active research only relatively recently~\cite{Asenjo2017,kornovan2019extremely,Chang2019,Albrecht2019,Molmer2019,Ke2019,Masson2020}. In particular, when the excitation fill factor $f=k/N$ is small,  subradiant states are antisymmetric products of single-particle subradiant states~\cite{Molmer2019,Zhong3phot2021}, reflecting so-called fermionization of atomic excitations.  
There also exists  an ``electron-hole'' symmetry between the fill factors $f$ and $1-f$. Based on such  symmetry one can expect interesting effects at the transition point  $f=1-f=1/2$ when the excitation degeneracy is at maximum. Indeed, many-body delocalization transition has been predicted  in disordered arrays for $f=1/2$, that can be also naively understood as suppression of disorder by electromagnetically induced transparency (EIT)~\cite{fayard2021manybody}. Many-body signatures in spontaneous emission cascade for atoms without a waveguide were revealed in \cite{Masson2020}. Despite this significant amount of recent theoretical research, to the best of our knowledge, subradiant multiple excited states in waveguide-coupled atomic arrays have  been directly probed only in one very recent experiment with just $N=4$ superconducting qubits~\cite{zanner2021coherent}. Moreover, their structure in the strongly many-body regime of $f\sim 1/2$, when the fermionic ansatz \cite{Molmer2019} is no longer valid, remains unclear.
\begin{figure}[b]
\centering\includegraphics[width=0.4\textwidth]{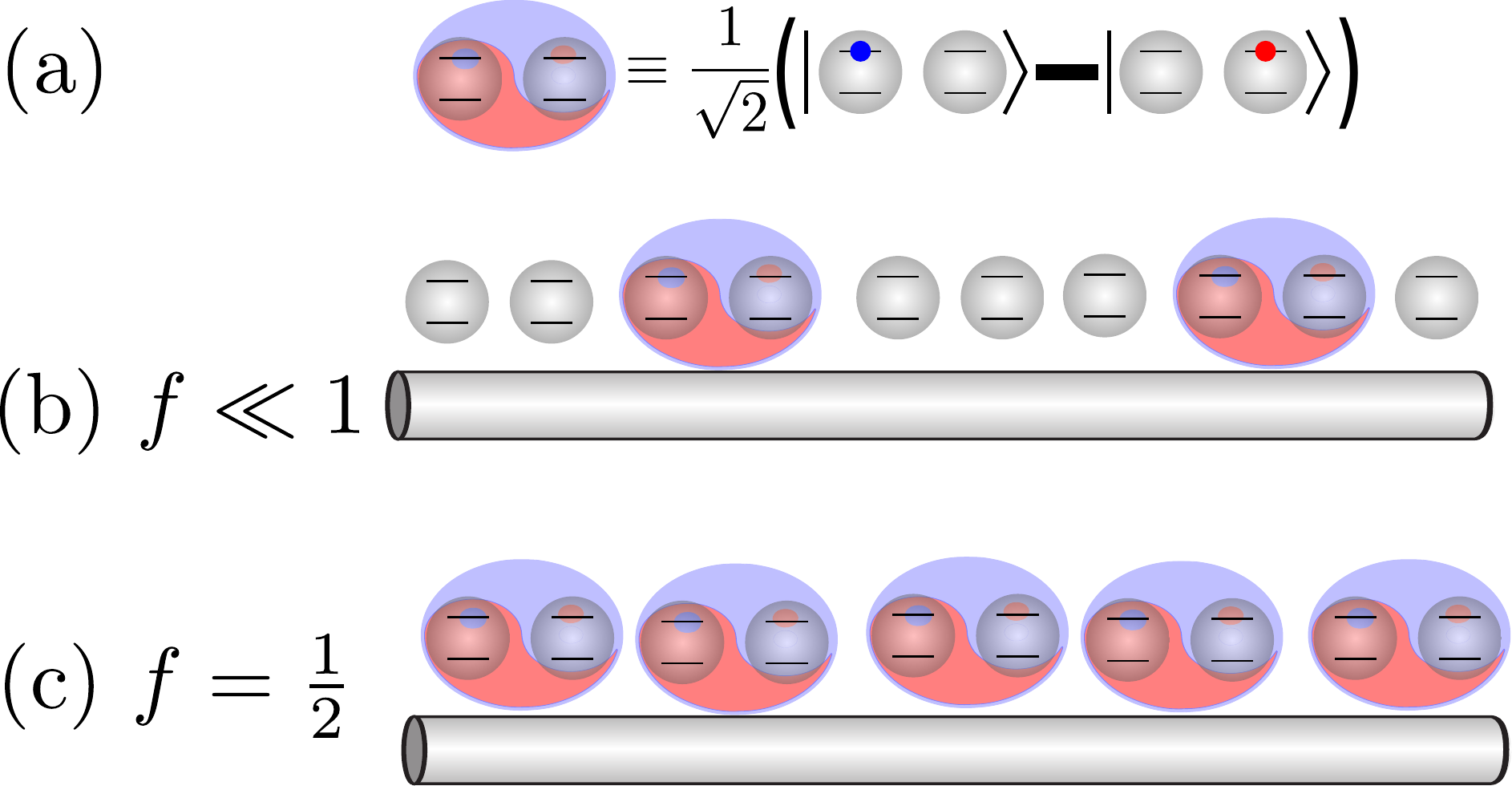}
\caption{Schematics  of (a)   single-excited dimer dark state for two atoms
(b)  double-excited dark state in array of atoms near a waveguide for fill-factor $f\ll 1$ and period $d\ll\lambda_0$.
(c)  many-body dark state for $f=1/2$.
}\label{fig:1}
\end{figure}

Here we focus on the subradiant states at large excitation fill factors and demonstrate a strong modification of their lifetimes and spatial structure at the transition point $f=1/2$.
Figure~\ref{fig:1} presents a simple qualitative picture behind this effect, that is valid in the limit of vanishing distance between the neighboring atoms, $d\to 0$. Namely, panel (a) illustrates a typical single-excited dark dimer state, an antisymmetric superpositions of two excited atoms $|\psi\rangle=(\sigma_1^\dag-\sigma_2^\dag)|0\rangle/\sqrt{2}$ (here $\sigma_{1,2}^\dag$ are the atomic raising operators).  Figure~\ref{fig:1}(b) shows an array with several independent single-excited dark dimers. Clearly, if the fill factor  is much smaller than unity, the interaction between different dimers can be neglected and we obtain a multiple-excited dark state. However, the maximal amount of non-interacting dark  dimers that can be put in the array with $N$ atoms is limited by $k=N/2$ and the corresponding half-filled array is illustrated in Figure~\ref{fig:1}(c). Based on such a naive analysis we expect that (i) multiple-excited dark (and subradiant) states do not exist for $f>1/2$ and that (ii) their spatial structure is strongly modified for $f=1/2$ when multiple dimers are squeezed together and become localized. 

{\it Lifetimes of subradiant states.} We now proceed to a more detailed analysis of the many-body excitation spectrum.
The structure under consideration is characterized by the following effective Hamiltonian, valid in the usual Markovian  and rotating wave approximations~\cite{Caneva2015,Ke2019,sheremet2021waveguide},
$H=-\rmi \GO\sum_{n,m=1}^N\sigma_n^\dag \sigma^{\vphantom{\dag}}_m \e^{\rmi \phi |m-n|}\:,$
where the energy is counted from the atomic resonance $\hbar\omega_0$ and $\varphi=\omega_0d/c\equiv 2\pi d/\lambda_0$ is the phase gained by light travelling the distance $d$ between two neighbouring atoms. The parameter $\GO\equiv \Gamma_{\rm 1D}/2$ is the radiative decay rate of  single atom into the waveguide, rendering the  effective Hamiltonian  non-Hermitian.
We are interested in the decay rates of  multiply excited subradiant eigenstates with $k$ excitations,
\begin{equation}
|\psi^{(k)}	\rangle=\sum\limits_{n_1n_2\ldots n_k=1}^N \psi_{n_1n_2\ldots n_k}\sigma_{n_1}^\dag \sigma_{n_2}^\dag \ldots \sigma_{n_k}^\dag |0\rangle\:.
\end{equation}
The decay rates are found numerically from the effective Schr\"odinger equation $H|\psi^{(k)}\rangle =k\eps |\psi ^{(k)}\rangle $ as $\Gamma=-\Im \eps$. 

\begin{figure}[t]
\centering\includegraphics[width=0.5\textwidth]{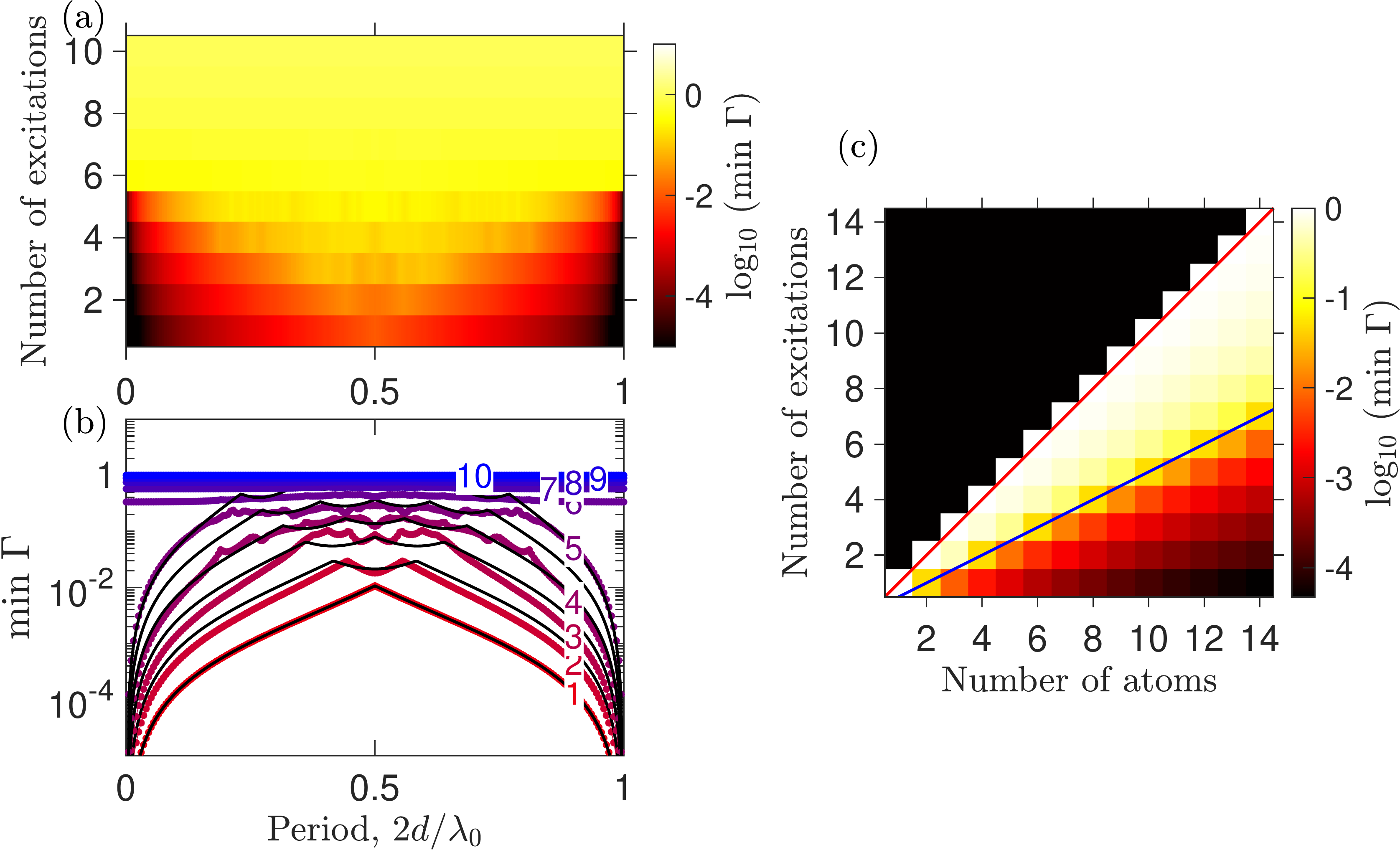}
\caption{(a),(b) Dependence of the radiative decay rate of the most-subradiant state $\rm min~\Gamma$ in the {$10$}-atom array on the array period $d$ and number of excitations $k$.
For (b) the number of excitations is shown near each curve. (c) Dependence of $\rm min~\Gamma$ on number of atoms $N$ and number of excitations $k$ calculated for a fixed period $d=0.05\lambda_0$.
Red and blue lines show the dependences $k=N$ (fill factor $f=1$) and $k=N/2$ ($f=1/2$), respectively. Decay rates are normalized to $\GO$. 
}\label{fig:2}
\end{figure}

Figures~\ref{fig:2}(a,c) present the dependence of the radiative decay rates of most subradiant states on the array period $d$, and the number of excitations $k$ for a fixed array size of $N=10$ atoms. The calculated decay rate is the smallest  when the period is close to $0$ or to $\lambda_0/2$. For $f \ll 1$, it scales as $\Gamma/\gamma_{\rm 1D}
\sim (d/\lambda_0)^2/N^3$ for $d\to 0$~\cite{Molmer2019}. For the excitation numbers $k$ smaller than $N/2$, the dependence of the decay rate on the array period  is qualitatively the same and  can be described by approximating  the multiple-excited subradiant states with antisymmetric combinations of most subradiant single-excited states with the decay rates $\Gamma^{(1)}_\nu$ \cite{Molmer2019},
$\Gamma^{(k)}=\sum_{\nu=1}^k\Gamma^{(1)}_\nu\:$ (black curves in Fig.~\ref{fig:2}c). 
However, the situation changes dramatically for $k>N/2$, when the excitation fill factor becomes larger than $1/2$. In this case, subradiant states disappear: all the decay rates become larger than those of a single atom, $\Gamma\gtrsim\gamma_{\rm 1D}$, in agreement with the qualitative picture in Fig.~\ref{fig:1}. Such behaviour is universal  as shown by the phase diagram Fig.~\ref{fig:2}(b) where we plot the decay rates of most subradiant states for a fixed period $d=0.05\lambda_0$ depending on both the number of atoms and the number of excitations. Red and blue lines correspond to the fully-excited and half-excited arrays, with fill factors $f=1$ and $f=1/2$. Clearly, the decay rates in the region between blue and red lines, where $f>1/2$, are significantly  larger than those for $f<1/2$.

The absence of subradiant states also follows from a general combinatoric argument: the  wavefunction with $k$ excitations is defined by $C_{N}^{k}$ complex amplitudes. In order to make the state subradiant one has to forbid its spontaneous decay into all the $C_{N}^{k-1}$ states with $k-1$ excitations. As soon as $k>N/2$, the number of conditions exceeds the number of independent amplitudes, $C_{N}^{k}>C_{N}^{k-1}$, and the dark states are ruled out.
\begin{figure*}[t]
\centering\includegraphics[width=\textwidth]{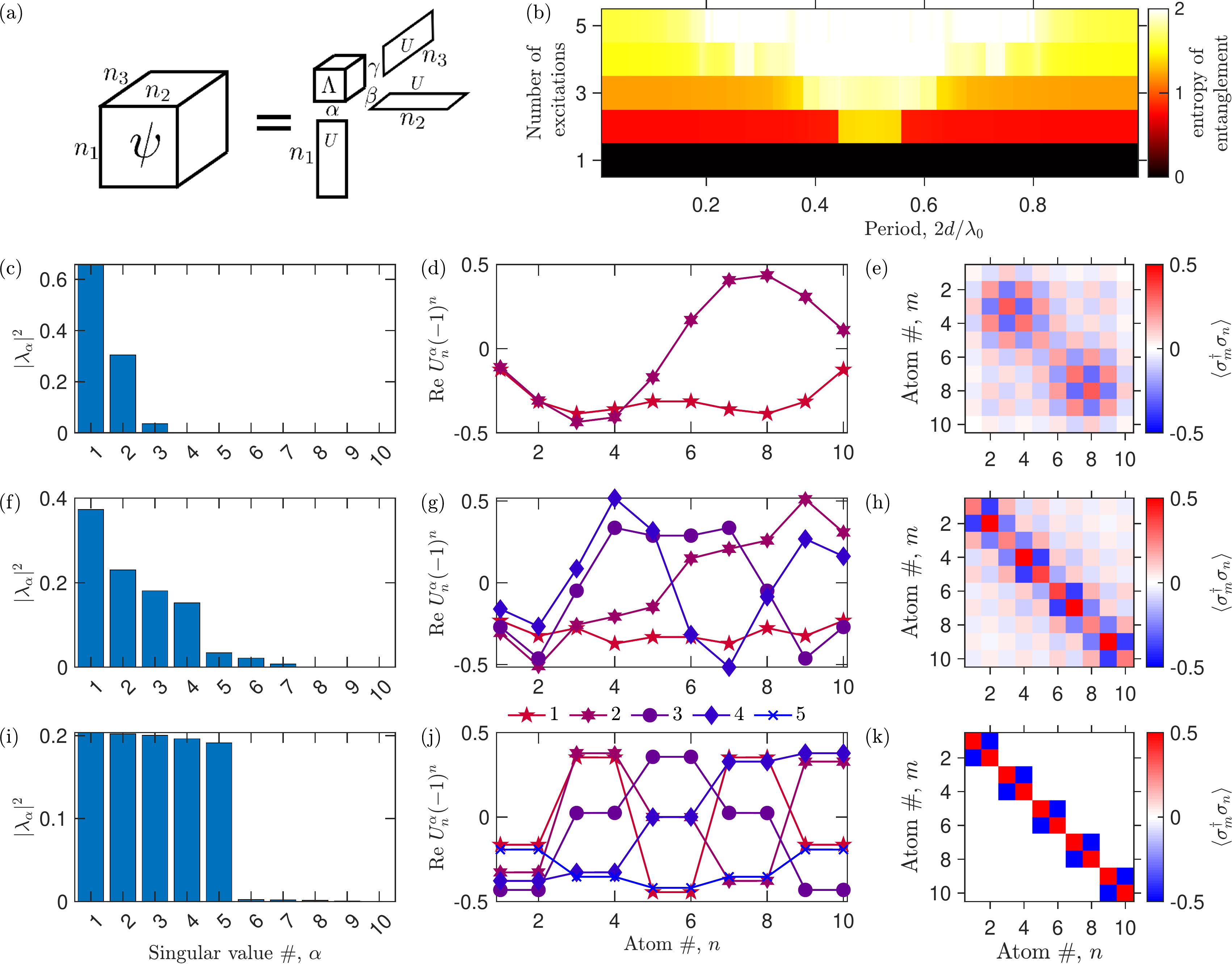}
\caption{(a) Illustration of high-order singular value decomposition for a three-particle wave function $\psi_{n_1n_2 n_3}$. (b) Entanglement entropy Eq.~\eqref{eq:entropy} calculated depending on the 
number of excitations $k$ and the array period  $d$ for $N=10$ atoms.
(c--k) Distributions of higher order singular values $\lambda_\alpha$ (left column), eigenvectors $U_n^\alpha$ (middle column),
and correlation functions $\langle\sigma_m^\dag \sigma_n\rangle$ (right column) 
 for the most subradiant states with
 two (c--e) ,three (f--h),and five (i--k) excitations. Calculation has been performed for $d=0.05\lambda_0$.
}\label{fig:3}
\end{figure*}

{\it Decompositon  of multiple-excited state over single-excited states.}
Next, we analyze the subradiant eigenstates in more detail. Due to the large size of the Hilbert space, even visualization of the numerically calculated wave function for $k\ge 3$ excitations is quite challenging~\cite{Zhong3phot2021}.
However, it is possible to use variational approximations to the full wave function, that represents the full $k$-rank tensor
$\psi_{n_1n_2\ldots n_k}$ as a product of several tensors of lower rank, such as matrix product states and tensor network technique
\cite{Schollwock2011,Orus2014}. Such approaches have already been succesfully applied to the setup of waveguide quantum electrodynamics~\cite{Chang2019,Regidor2021}. Here, we use a slightly different technique of multilinear singular value decomposition
\cite{DeLathauwer2000,tensorlab}, that, contrary to the matrix product state approach, is numerically exact. In this method the $k$-rank symmetric tensor $\psi_{lm\ldots n}$ is factorized as
\begin{equation}
\psi_{n_1n_2\ldots n_k}= \sum_{\alpha_1\alpha_2\ldots \alpha_k=1}^N\Lambda_{\alpha_1\alpha_2\ldots \alpha_k}U^{\alpha_1}_{n_1} U^{\alpha_2}_{n_2}\ldots U^{\alpha_k}_{n_k}\:,\label{eq:hosvd}
\end{equation}
where the tensor  $U$ is unitary, $\sum_{n=1}^N {U^\alpha_n}^* U^{\alpha'}_n=\delta_{\alpha\alpha'} $,  and the so-called core  tensor $\Lambda$ is symmetric and quasi-diagonal, i.e., it  satisfies the identity
$\sum_{\alpha_2\ldots \alpha_k=1}^N\Lambda^{*}_{\alpha_1\alpha_2\ldots \alpha_k}\Lambda_{\alpha_1'\alpha_2\ldots \alpha_k}=0$\text{ for }$\alpha_1\ne \alpha_1'$.
The decomposition Eq.~\eqref{eq:hosvd} can be viewed as a generalization of the conventional Schmidt (singular) value decomposition of a two-particle wave function and it is schematically illustrated in Fig.~\ref{fig:3}(a).
In the multilinear decomposition, the role of singular values is played by the Frobenius norms of the subtensor of the  tensor $\Lambda$ defined as \cite{DeLathauwer2000}
\[
\lambda_\alpha={\rm frob}\:\Lambda_{\alpha,:} \equiv \sqrt{\sum_{\alpha_2\ldots \alpha_k=1}^N |\Lambda_{\alpha,\alpha_2\ldots \alpha_k}|^2}\:.
\] Due to the orthogonality properties of the tensors $\Lambda$ and $U$, the sum of the components of the $|\psi_{lm\ldots n}|^2$ over all indices equals just to 
$\sum_{\alpha=1}^N|\lambda_\alpha|^2$, similarly to the case of usual single-particle Schmidt decomposition. This analogy allows us to introduce  the generalized entropy of entanglement \cite{Eisert2010} as
\begin{equation}\label{eq:entropy}
S=-\sum\limits_{\alpha=1}^N |\lambda_\alpha|^2 \ln |\lambda_\alpha|^2\:.
\end{equation}
Equation~\eqref{eq:entropy} approximately quantifies  the number of single-particle states $U^\alpha$  necessary to describe a $k$-excited state $\psi$.
We note that such decomposition is efficient only for $k\le N/2$. For larger $k$ values, it is more instructive to exploit the ``electron-hole symmetry'' of the problem \cite{Chang2019} and represent the $k$ atomic excitations as $N-k$ ``holes'' in the array of fully excited atoms.

 The dependence of entanglement entropy of  most subradiant states on the number of excitations and the array period  is shown in Fig.~\ref{fig:3}(b). These results are in good qualitative agreement with the radiative decay rates in Fig.~\ref{fig:2}. The entropy increases, i.e. the states become more complex, for (i) larger numbers of excitations and (ii) stronger detuning of the period $d$   from the degeneracy points $d=0$ and $d=\lambda_0/2$. We show explicitly all the multilinear singular values in Fig.~\ref{fig:3}(c,f,i). The  panels from top to bottom correspond to increasing numbers of excitations for a given period $d=0.05\lambda_0$. The general observation is that a subradiant state with $k$ excitations has first $k$  multilinear singular values much larger than the remaining $N-k$ ones. This confirms our qualitative interpretation of Eq.~\eqref{eq:hosvd} as the expansion of a many-body state over single-particle ones. Crucially, the  obtained single-particle states $U^\alpha$ strongly depend on the number of excitations, i.e. they are renormalized by interactions. When the fill factor is small, as in the case of $f=1/5$ in   Fig.~\ref{fig:3}(d), the states $U^{1,2}$ are just two standing waves with zero and one node, in agreement with the fermionic ansatz of Ref.~\cite{Molmer2019}:
 \begin{equation}
 U_n^{1}\propto (-1)^n \sin \frac{\pi n}{N},\quad 
  U_n^{2}\propto (-1)^n \sin \frac{2\pi n}{N},\:n=1\ldots N\label{eq:ferm}\:.
 \end{equation} 
 However, the functions $U_n^\alpha$ drastically change at the threshold, for $k=N/2$ excitations, see Fig.~\ref{fig:3}(j). 
A careful inspection reveals that the functions $U_n^\alpha$ become ``dimerized'', i.e, have equal amplitudes (up to the sign) at neighbouring sites $2k-1$ and $2k$, supporting the naive picture Fig.~\ref{fig:1}(c). More specifically, the interaction of the dimers leads to formation of the ``dimerized'' standing waves
  \begin{equation}
 U_{2k-1}^{\alpha}=-U_{2k}^{\alpha}\propto \cos \frac{2k\pi  \alpha}{N},\quad k=0,1\ldots \frac{N}{2}\label{eq:ferm2}\:,
 \end{equation} 
 where the ``elementary unit'' is twice larger than for the normal standing wave in Eq.~\eqref{eq:ferm}.  
Somewhat similar spontaneously symmetry breaking transitions accompanied by the  doubling of the unit cell size, are known for classical nonlinear equations and have been experimentally observed  for interacting excitonic-polaritons in one-dimensional waveguides~\cite{Zhang2015,Nalitov2017}.
 The onset of  dimerization effect Eq.~\eqref{eq:ferm2} can also  be seen already for a double-excited subradiant state in a four-atom array, that reads $|\psi\rangle=\frac{1}{2}(\sigma_1^\dag-\sigma_2^\dag)(\sigma_3^\dag-\sigma_4^\dag)|0\rangle$ for $d\ll \lambda_0$~\cite{Ke2019}. However, since in 
Ref.~\cite{Ke2019} we have restricted ourselves just to the double-excited states, we were not able to resolve the difference between the decompositions Eq.~\eqref{eq:ferm} and Eq.~\eqref{eq:ferm2}  that has  become evident only in the many-atom-many-excitation regime of Fig.~\ref{fig:3}. 

  The  dimerization of subradiant states is directly visualized by their spin-spin correlation function $\langle\sigma_m^\dag \sigma_n\rangle$, plotted in the right column of Fig.~\ref{fig:3}. For $k=2$ excitations the correlations are long-ranged, reflecting the spatial profile of the eigenstates Eq.~\eqref{eq:ferm}, see Fig.~\ref{fig:3}e. At the threshold, for $k=N/2$, the correlations are increased and become short-ranged, see Fig.~\ref{fig:3}k. The only significant elements of the correlation matrix at the threshold are
 \begin{equation}\label{eq:corr}
  \langle\sigma_m^\dag \sigma_m^{\vphantom{\dag}}\rangle\approx \frac{1}{2} \text{ and }\langle\sigma_{2j-1}^\dag \sigma_{2j}^{\vphantom{\dag}}\rangle\approx -\frac{1}{2}\:,
  \end{equation}
  i.e., there appears a short-range effective antiferromagnetic order.

\begin{figure}[t]
\centering\includegraphics[width=0.48\textwidth]{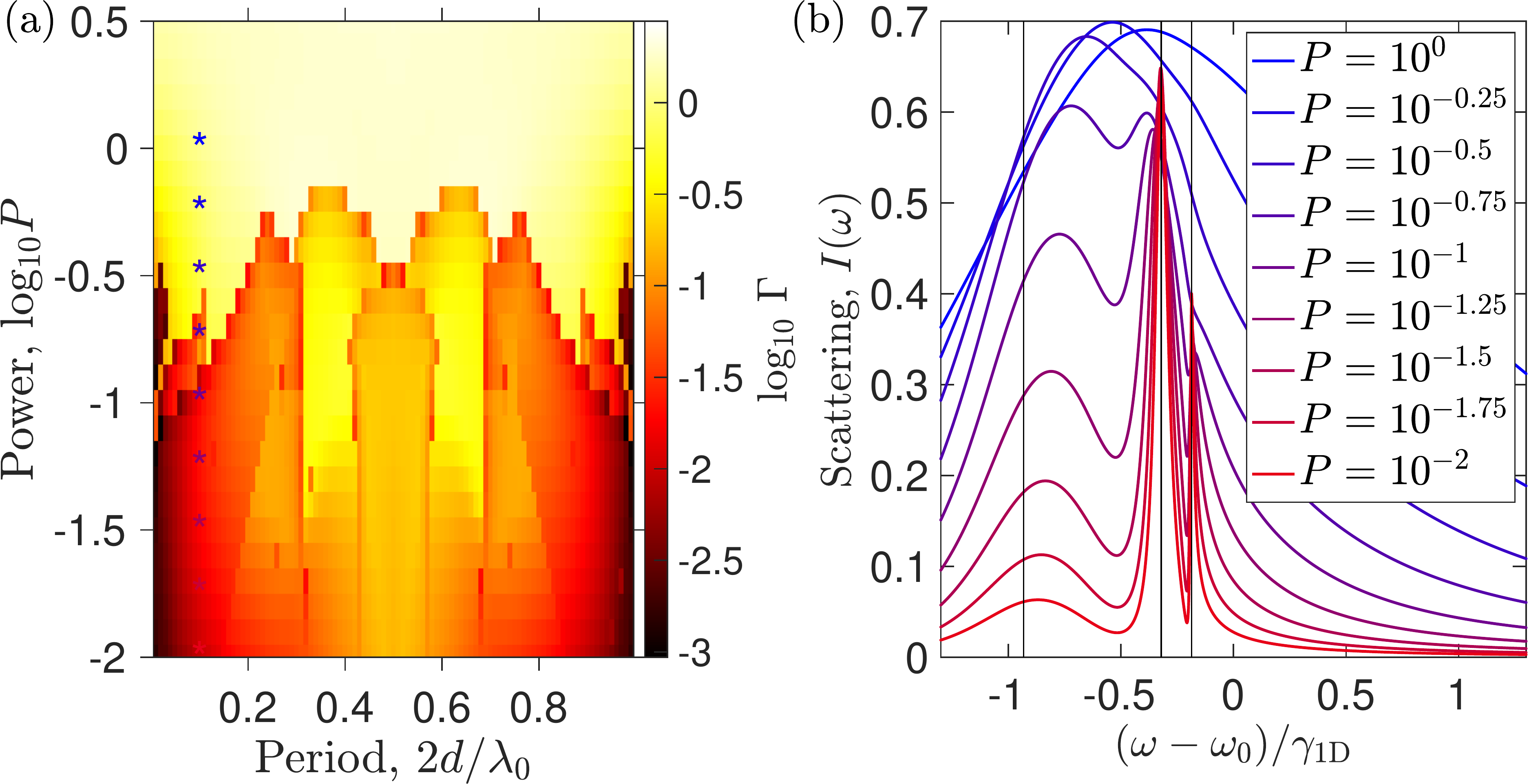}
\caption{(a) Linewidth of narrowest resonance feature in the incoherent scattering spectra depending on the pump power $P$ and the array period.
(b) Incoherent scattering spectra calculated for an array of $N=4$~atoms with the period $d=0.05\lambda_0$ for several powers indicated on graph.
Stars  in (a) indicate the values of parameters used for calculation in (b). Vertical lines show the positions of two most subradiant states.
 The excitation power has been normalized to $\gamma_{\rm 1D}$. 
}\label{fig:4}
\end{figure}

{\it Detection of the subradiant-to-bright transition.} Experimental characterization of multiple-excited subradiant states remains an important standing problem. 
While several single-excited subradiant states have been successfully observed as reflection resonances for an array of 8 superconducting qubits~\cite{brehm2020waveguide}, even for double-excited subradiant states to the best of our knowledge  there exists only one preliminary  observation \cite{zanner2021coherent} , that has been peformed for two pairs of coupled qubits, rather than for an equidistant array.  Since these states are weakly coupled to the waveguide modes, it might be more efficient to excite them locally instead, by selectively pumping individual atoms or qubits and measuring the spin-spin correlation functions depending on power to verify Eq.~\eqref{eq:corr}.  However, interesting results can be potentially obtained even from the incoherent scattering spectra measured directly through the waveguide.  We assume that the array is excited from the left by a coherent field with the frequency $\omega$, that is described by the coupling term
$(\sqrt{P}/\gamma_{\rm 1D})\sum_{j=1}^N (\sigma_j \e^{-\rmi (\omega-\omega_0)t}+{\rm H.c.})$
in the electric dipole and rotating wave approximations, where $P$ is the normalized input power. Next,  we use the  input-output theory \cite{Blais2013} and calculate the total amount of incoherently scattered photons $I(\omega)=1-|r(\omega)|^2-|t(\omega)|^2$, where  $r(\omega)$ and $t(\omega)$ are the amplitudes of coherent reflection and transmission coefficients. Examples of incoherent scattering spectra  calculated for a small $4$-atom array with a fixed period $d=0.05\lambda_0$ are shown in Fig.~\ref{fig:4}(b). At low power, they show narrow peaks centered  at the frequencies of the single-excited subradiant states, shown  by vertical lines in Fig.~\ref{fig:4}(b). Increase of the power leads to the spectral broadening and also appearance of new peaks in the spectrum.    Next, we extract from each  incoherent scattering spectra the linewidth of the narrowest peak and plot it by color in 
Fig.~\ref{fig:4} as a function the pump power $P$ and the array period. Qualitatively, the result is quite similar to those in Fig.~\ref{fig:2}a: narrow spectral features, corresponding to subradiant states, become wider with the power increase and  disappear above a certain threshold.  Similarly to Fig.~\ref{fig:2}a, an interesting non-monotonous behaviour is observed near the anti-Bragg period $d=\lambda_0/4$. However, the observed spectral broadening in Fig.~\ref{fig:4} is caused not only by the excitation of many-particle states, but also by the pump-induced broadening of the transitions.
Moreover, we must keep in mind that when the array is excited through the waveguide the occupation numbers $\langle\sigma_m^\dag \sigma_m\rangle$ stay below $1/2$ for all the qubits  and the threshold $f=1/2$ of the half-excited array is never crossed.  Despite this limitations, we believe that observation of the evolution of the incoherent scattering spectra with pump power similar to Fig.~\ref{fig:4} can be an important precursor of  the subradiant-to-bright transition. 

{\it Outlook.} Our findings provide yet another demonstration of the fundamental many-body physics in the waveguide quantum electrodynamics setup. In this work, we have limited ourselves to relatively short arrays with just $N\lesssim 10$ atoms, that are well within the range of state-of-the-art experimental structures with superconducting qubits~\cite{Ye2019}. We expect much richer physics for larger arrays, when  bound photon pairs  start playing role~\cite{Zhang2019arXiv}. For example, it is quite intriguing whether ``magic periods'' such as $d=\lambda_0/12$ \cite{Poddubny2019quasiflat,Zhang2020d} with quasi-flat band of composite excitations survive in the many-body regime. Another standing problem is the influence of disorder and the possible interplay of the many-body delocalization transition \cite{fayard2021manybody} with the subradiant-to-bright transition for $f=1/2$ fill factor. On the more applied side, our results could be useful to design long-living complex quantum correlations.

\acknowledgements
The authors are grateful to M.M.~Glazov and M.O.~Nestoklon for useful discussions.


\begin{thebibliography}{32}%
\makeatletter
\providecommand \@ifxundefined [1]{%
 \@ifx{#1\undefined}
}%
\providecommand \@ifnum [1]{%
 \ifnum #1\expandafter \@firstoftwo
 \else \expandafter \@secondoftwo
 \fi
}%
\providecommand \@ifx [1]{%
 \ifx #1\expandafter \@firstoftwo
 \else \expandafter \@secondoftwo
 \fi
}%
\providecommand \natexlab [1]{#1}%
\providecommand \enquote  [1]{``#1''}%
\providecommand \bibnamefont  [1]{#1}%
\providecommand \bibfnamefont [1]{#1}%
\providecommand \citenamefont [1]{#1}%
\providecommand \href@noop [0]{\@secondoftwo}%
\providecommand \href [0]{\begingroup \@sanitize@url \@href}%
\providecommand \@href[1]{\@@startlink{#1}\@@href}%
\providecommand \@@href[1]{\endgroup#1\@@endlink}%
\providecommand \@sanitize@url [0]{\catcode `\\12\catcode `\$12\catcode
  `\&12\catcode `\#12\catcode `\^12\catcode `\_12\catcode `\%12\relax}%
\providecommand \@@startlink[1]{}%
\providecommand \@@endlink[0]{}%
\providecommand \url  [0]{\begingroup\@sanitize@url \@url }%
\providecommand \@url [1]{\endgroup\@href {#1}{\urlprefix }}%
\providecommand \urlprefix  [0]{URL }%
\providecommand \Eprint [0]{\href }%
\providecommand \doibase [0]{http://dx.doi.org/}%
\providecommand \selectlanguage [0]{\@gobble}%
\providecommand \bibinfo  [0]{\@secondoftwo}%
\providecommand \bibfield  [0]{\@secondoftwo}%
\providecommand \translation [1]{[#1]}%
\providecommand \BibitemOpen [0]{}%
\providecommand \bibitemStop [0]{}%
\providecommand \bibitemNoStop [0]{.\EOS\space}%
\providecommand \EOS [0]{\spacefactor3000\relax}%
\providecommand \BibitemShut  [1]{\csname bibitem#1\endcsname}%
\let\auto@bib@innerbib\@empty
\bibitem [{\citenamefont {Dicke}(1954)}]{Dicke1954}%
  \BibitemOpen
  \bibfield  {author} {\bibinfo {author} {\bibfnamefont {R.~H.}\ \bibnamefont
  {Dicke}},\ }\bibfield  {title} {\enquote {\bibinfo {title} {{C}oherence in
  {S}pontaneous {R}adiation {P}rocesses},}\ }\href {\doibase
  10.1103/PhysRev.93.99} {\bibfield  {journal} {\bibinfo  {journal} {Phys.
  Rev.}\ }\textbf {\bibinfo {volume} {93}},\ \bibinfo {pages} {99} (\bibinfo
  {year} {1954})}\BibitemShut {NoStop}%
\bibitem [{\citenamefont {Scully}\ and\ \citenamefont
  {Svidzinsky}(2009)}]{Scully2009}%
  \BibitemOpen
  \bibfield  {author} {\bibinfo {author} {\bibfnamefont {M.~O.}\ \bibnamefont
  {Scully}}\ and\ \bibinfo {author} {\bibfnamefont {A.~A.}\ \bibnamefont
  {Svidzinsky}},\ }\bibfield  {title} {\enquote {\bibinfo {title} {The super of
  superradiance},}\ }\href {\doibase 10.1126/science.1176695} {\bibfield
  {journal} {\bibinfo  {journal} {Science}\ }\textbf {\bibinfo {volume}
  {325}},\ \bibinfo {pages} {1510--1511} (\bibinfo {year} {2009})}\BibitemShut
  {NoStop}%
\bibitem [{\citenamefont {Roy}\ \emph {et~al.}(2017)\citenamefont {Roy},
  \citenamefont {Wilson},\ and\ \citenamefont {Firstenberg}}]{Roy2017}%
  \BibitemOpen
  \bibfield  {author} {\bibinfo {author} {\bibfnamefont {D.}~\bibnamefont
  {Roy}}, \bibinfo {author} {\bibfnamefont {C.~M.}\ \bibnamefont {Wilson}}, \
  and\ \bibinfo {author} {\bibfnamefont {O.}~\bibnamefont {Firstenberg}},\
  }\bibfield  {title} {\enquote {\bibinfo {title} {\textit{Colloquium:}
  strongly interacting photons in one-dimensional continuum},}\ }\href
  {\doibase 10.1103/RevModPhys.89.021001} {\bibfield  {journal} {\bibinfo
  {journal} {Rev. Mod. Phys.}\ }\textbf {\bibinfo {volume} {89}},\ \bibinfo
  {pages} {021001} (\bibinfo {year} {2017})}\BibitemShut {NoStop}%
\bibitem [{\citenamefont {Chang}\ \emph {et~al.}(2018)\citenamefont {Chang},
  \citenamefont {Douglas}, \citenamefont {Gonz\'alez-Tudela}, \citenamefont
  {Hung},\ and\ \citenamefont {Kimble}}]{KimbleRMP2018}%
  \BibitemOpen
  \bibfield  {author} {\bibinfo {author} {\bibfnamefont {D.~E.}\ \bibnamefont
  {Chang}}, \bibinfo {author} {\bibfnamefont {J.~S.}\ \bibnamefont {Douglas}},
  \bibinfo {author} {\bibfnamefont {A.}~\bibnamefont {Gonz\'alez-Tudela}},
  \bibinfo {author} {\bibfnamefont {C.-L.}\ \bibnamefont {Hung}}, \ and\
  \bibinfo {author} {\bibfnamefont {H.~J.}\ \bibnamefont {Kimble}},\ }\bibfield
   {title} {\enquote {\bibinfo {title} {\textit{Colloquium:} quantum matter
  built from nanoscopic lattices of atoms and photons},}\ }\href {\doibase
  10.1103/RevModPhys.90.031002} {\bibfield  {journal} {\bibinfo  {journal}
  {Rev. Mod. Phys.}\ }\textbf {\bibinfo {volume} {90}},\ \bibinfo {pages}
  {031002} (\bibinfo {year} {2018})}\BibitemShut {NoStop}%
\bibitem [{\citenamefont {Sheremet}\ \emph {et~al.}(2021)\citenamefont
  {Sheremet}, \citenamefont {Petrov}, \citenamefont {Iorsh}, \citenamefont
  {Poshakinskiy},\ and\ \citenamefont {Poddubny}}]{sheremet2021waveguide}%
  \BibitemOpen
  \bibfield  {author} {\bibinfo {author} {\bibfnamefont {A.~S.}\ \bibnamefont
  {Sheremet}}, \bibinfo {author} {\bibfnamefont {M.~I.}\ \bibnamefont
  {Petrov}}, \bibinfo {author} {\bibfnamefont {I.~V.}\ \bibnamefont {Iorsh}},
  \bibinfo {author} {\bibfnamefont {A.~V.}\ \bibnamefont {Poshakinskiy}}, \
  and\ \bibinfo {author} {\bibfnamefont {A.~N.}\ \bibnamefont {Poddubny}},\
  }\href@noop {} {\enquote {\bibinfo {title} {Waveguide quantum
  electrodynamics: collective radiance and photon-photon correlations},}\ }
  (\bibinfo {year} {2021}),\ \Eprint {http://arxiv.org/abs/2103.06824}
  {arXiv:2103.06824 [quant-ph]} \BibitemShut {NoStop}%
\bibitem [{\citenamefont {MacGillivray}\ and\ \citenamefont
  {Feld}(1976)}]{Feld1976}%
  \BibitemOpen
  \bibfield  {author} {\bibinfo {author} {\bibfnamefont {J.~C.}\ \bibnamefont
  {MacGillivray}}\ and\ \bibinfo {author} {\bibfnamefont {M.~S.}\ \bibnamefont
  {Feld}},\ }\bibfield  {title} {\enquote {\bibinfo {title} {Theory of
  superradiance in an extended, optically thick medium},}\ }\href {\doibase
  10.1103/PhysRevA.14.1169} {\bibfield  {journal} {\bibinfo  {journal} {Phys.
  Rev. A}\ }\textbf {\bibinfo {volume} {14}},\ \bibinfo {pages} {1169--1189}
  (\bibinfo {year} {1976})}\BibitemShut {NoStop}%
\bibitem [{\citenamefont {Yudson}\ and\ \citenamefont
  {Rupasov}(1984)}]{Yudson1984}%
  \BibitemOpen
  \bibfield  {author} {\bibinfo {author} {\bibfnamefont {V.}~\bibnamefont
  {Yudson}}\ and\ \bibinfo {author} {\bibfnamefont {V.}~\bibnamefont
  {Rupasov}},\ }\bibfield  {title} {\enquote {\bibinfo {title} {Exact {Dicke}
  superradiance theory: Bethe wavefunctions in the discrete atom model},}\
  }\href {http://www.jetp.ac.ru/cgi-bin/e/index/e/59/3/p478?a=list} {\bibfield
  {journal} {\bibinfo  {journal} {Sov. Phys. JETP}\ }\textbf {\bibinfo {volume}
  {59}},\ \bibinfo {pages} {478} (\bibinfo {year} {1984})}\BibitemShut
  {NoStop}%
\bibitem [{\citenamefont {Henriet}\ \emph {et~al.}(2019)\citenamefont
  {Henriet}, \citenamefont {Douglas}, \citenamefont {Chang},\ and\
  \citenamefont {Albrecht}}]{Chang2019}%
  \BibitemOpen
  \bibfield  {author} {\bibinfo {author} {\bibfnamefont {L.}~\bibnamefont
  {Henriet}}, \bibinfo {author} {\bibfnamefont {J.~S.}\ \bibnamefont
  {Douglas}}, \bibinfo {author} {\bibfnamefont {D.~E.}\ \bibnamefont {Chang}},
  \ and\ \bibinfo {author} {\bibfnamefont {A.}~\bibnamefont {Albrecht}},\
  }\bibfield  {title} {\enquote {\bibinfo {title} {Critical open-system
  dynamics in a one-dimensional optical-lattice clock},}\ }\href {\doibase
  10.1103/PhysRevA.99.023802} {\bibfield  {journal} {\bibinfo  {journal} {Phys.
  Rev. A}\ }\textbf {\bibinfo {volume} {99}},\ \bibinfo {pages} {023802}
  (\bibinfo {year} {2019})}\BibitemShut {NoStop}%
\bibitem [{\citenamefont {Masson}\ \emph {et~al.}(2020)\citenamefont {Masson},
  \citenamefont {Ferrier-Barbut}, \citenamefont {Orozco}, \citenamefont
  {Browaeys},\ and\ \citenamefont {Asenjo-Garcia}}]{Masson2020}%
  \BibitemOpen
  \bibfield  {author} {\bibinfo {author} {\bibfnamefont {S.~J.}\ \bibnamefont
  {Masson}}, \bibinfo {author} {\bibfnamefont {I.}~\bibnamefont
  {Ferrier-Barbut}}, \bibinfo {author} {\bibfnamefont {L.~A.}\ \bibnamefont
  {Orozco}}, \bibinfo {author} {\bibfnamefont {A.}~\bibnamefont {Browaeys}}, \
  and\ \bibinfo {author} {\bibfnamefont {A.}~\bibnamefont {Asenjo-Garcia}},\
  }\bibfield  {title} {\enquote {\bibinfo {title} {Many-body signatures of
  collective decay in atomic chains},}\ }\href {\doibase
  10.1103/PhysRevLett.125.263601} {\bibfield  {journal} {\bibinfo  {journal}
  {Phys. Rev. Lett.}\ }\textbf {\bibinfo {volume} {125}},\ \bibinfo {pages}
  {263601} (\bibinfo {year} {2020})}\BibitemShut {NoStop}%
\bibitem [{\citenamefont {Asenjo-Garcia}\ \emph {et~al.}(2017)\citenamefont
  {Asenjo-Garcia}, \citenamefont {Moreno-Cardoner}, \citenamefont {Albrecht},
  \citenamefont {Kimble},\ and\ \citenamefont {Chang}}]{Asenjo2017}%
  \BibitemOpen
  \bibfield  {author} {\bibinfo {author} {\bibfnamefont {A.}~\bibnamefont
  {Asenjo-Garcia}}, \bibinfo {author} {\bibfnamefont {M.}~\bibnamefont
  {Moreno-Cardoner}}, \bibinfo {author} {\bibfnamefont {A.}~\bibnamefont
  {Albrecht}}, \bibinfo {author} {\bibfnamefont {H.~J.}\ \bibnamefont
  {Kimble}}, \ and\ \bibinfo {author} {\bibfnamefont {D.~E.}\ \bibnamefont
  {Chang}},\ }\bibfield  {title} {\enquote {\bibinfo {title} {Exponential
  improvement in photon storage fidelities using subradiance and ``selective
  radiance'' in atomic arrays},}\ }\href {\doibase 10.1103/PhysRevX.7.031024}
  {\bibfield  {journal} {\bibinfo  {journal} {Phys. Rev. X}\ }\textbf {\bibinfo
  {volume} {7}},\ \bibinfo {pages} {031024} (\bibinfo {year}
  {2017})}\BibitemShut {NoStop}%
\bibitem [{\citenamefont {Kornovan}\ \emph {et~al.}(2019)\citenamefont
  {Kornovan}, \citenamefont {Corzo}, \citenamefont {Laurat},\ and\
  \citenamefont {Sheremet}}]{kornovan2019extremely}%
  \BibitemOpen
  \bibfield  {author} {\bibinfo {author} {\bibfnamefont {D.~F.}\ \bibnamefont
  {Kornovan}}, \bibinfo {author} {\bibfnamefont {N.~V.}\ \bibnamefont {Corzo}},
  \bibinfo {author} {\bibfnamefont {J.}~\bibnamefont {Laurat}}, \ and\ \bibinfo
  {author} {\bibfnamefont {A.~S.}\ \bibnamefont {Sheremet}},\ }\bibfield
  {title} {\enquote {\bibinfo {title} {Extremely subradiant states in a
  periodic one-dimensional atomic array},}\ }\href {\doibase
  10.1103/PhysRevA.100.063832} {\bibfield  {journal} {\bibinfo  {journal}
  {Phys. Rev. A}\ }\textbf {\bibinfo {volume} {100}},\ \bibinfo {pages}
  {063832} (\bibinfo {year} {2019})}\BibitemShut {NoStop}%
\bibitem [{\citenamefont {Albrecht}\ \emph {et~al.}(2019)\citenamefont
  {Albrecht}, \citenamefont {Henriet}, \citenamefont {Asenjo-Garcia},
  \citenamefont {Dieterle}, \citenamefont {Painter},\ and\ \citenamefont
  {Chang}}]{Albrecht2019}%
  \BibitemOpen
  \bibfield  {author} {\bibinfo {author} {\bibfnamefont {A.}~\bibnamefont
  {Albrecht}}, \bibinfo {author} {\bibfnamefont {L.}~\bibnamefont {Henriet}},
  \bibinfo {author} {\bibfnamefont {A.}~\bibnamefont {Asenjo-Garcia}}, \bibinfo
  {author} {\bibfnamefont {P.~B.}\ \bibnamefont {Dieterle}}, \bibinfo {author}
  {\bibfnamefont {O.}~\bibnamefont {Painter}}, \ and\ \bibinfo {author}
  {\bibfnamefont {D.~E.}\ \bibnamefont {Chang}},\ }\bibfield  {title} {\enquote
  {\bibinfo {title} {Subradiant states of quantum bits coupled to a
  one-dimensional waveguide},}\ }\href {\doibase 10.1088/1367-2630/ab0134}
  {\bibfield  {journal} {\bibinfo  {journal} {New J. Phys.}\ }\textbf {\bibinfo
  {volume} {21}},\ \bibinfo {pages} {025003} (\bibinfo {year}
  {2019})}\BibitemShut {NoStop}%
\bibitem [{\citenamefont {Zhang}\ and\ \citenamefont
  {M\o{}lmer}(2019)}]{Molmer2019}%
  \BibitemOpen
  \bibfield  {author} {\bibinfo {author} {\bibfnamefont {Y.-X.}\ \bibnamefont
  {Zhang}}\ and\ \bibinfo {author} {\bibfnamefont {K.}~\bibnamefont
  {M\o{}lmer}},\ }\bibfield  {title} {\enquote {\bibinfo {title} {Theory of
  subradiant states of a one-dimensional two-level atom chain},}\ }\href
  {\doibase 10.1103/PhysRevLett.122.203605} {\bibfield  {journal} {\bibinfo
  {journal} {Phys. Rev. Lett.}\ }\textbf {\bibinfo {volume} {122}},\ \bibinfo
  {pages} {203605} (\bibinfo {year} {2019})}\BibitemShut {NoStop}%
\bibitem [{\citenamefont {Ke}\ \emph {et~al.}(2019)\citenamefont {Ke},
  \citenamefont {Poshakinskiy}, \citenamefont {Lee}, \citenamefont {Kivshar},\
  and\ \citenamefont {Poddubny}}]{Ke2019}%
  \BibitemOpen
  \bibfield  {author} {\bibinfo {author} {\bibfnamefont {Y.}~\bibnamefont
  {Ke}}, \bibinfo {author} {\bibfnamefont {A.~V.}\ \bibnamefont
  {Poshakinskiy}}, \bibinfo {author} {\bibfnamefont {C.}~\bibnamefont {Lee}},
  \bibinfo {author} {\bibfnamefont {Y.~S.}\ \bibnamefont {Kivshar}}, \ and\
  \bibinfo {author} {\bibfnamefont {A.~N.}\ \bibnamefont {Poddubny}},\
  }\bibfield  {title} {\enquote {\bibinfo {title} {Inelastic scattering of
  photon pairs in qubit arrays with subradiant states},}\ }\href {\doibase
  10.1103/PhysRevLett.123.253601} {\bibfield  {journal} {\bibinfo  {journal}
  {Phys. Rev. Lett.}\ }\textbf {\bibinfo {volume} {123}},\ \bibinfo {pages}
  {253601} (\bibinfo {year} {2019})}\BibitemShut {NoStop}%
\bibitem [{\citenamefont {Zhong}\ and\ \citenamefont
  {Poddubny}(2021)}]{Zhong3phot2021}%
  \BibitemOpen
  \bibfield  {author} {\bibinfo {author} {\bibfnamefont {J.}~\bibnamefont
  {Zhong}}\ and\ \bibinfo {author} {\bibfnamefont {A.~N.}\ \bibnamefont
  {Poddubny}},\ }\bibfield  {title} {\enquote {\bibinfo {title} {Classification
  of three-photon states in waveguide quantum electrodynamics},}\ }\href
  {\doibase 10.1103/PhysRevA.103.023720} {\bibfield  {journal} {\bibinfo
  {journal} {Phys. Rev. A}\ }\textbf {\bibinfo {volume} {103}},\ \bibinfo
  {pages} {023720} (\bibinfo {year} {2021})}\BibitemShut {NoStop}%
\bibitem [{\citenamefont {Fayard}\ \emph {et~al.}(2021)\citenamefont {Fayard},
  \citenamefont {Henriet}, \citenamefont {Asenjo-Garcia},\ and\ \citenamefont
  {Chang}}]{fayard2021manybody}%
  \BibitemOpen
  \bibfield  {author} {\bibinfo {author} {\bibfnamefont {N.}~\bibnamefont
  {Fayard}}, \bibinfo {author} {\bibfnamefont {L.}~\bibnamefont {Henriet}},
  \bibinfo {author} {\bibfnamefont {A.}~\bibnamefont {Asenjo-Garcia}}, \ and\
  \bibinfo {author} {\bibfnamefont {D.}~\bibnamefont {Chang}},\ }\href@noop {}
  {\enquote {\bibinfo {title} {Many-body localization in waveguide {QED}},}\ }
  (\bibinfo {year} {2021}),\ \Eprint {http://arxiv.org/abs/2101.01645}
  {arXiv:2101.01645 [quant-ph]} \BibitemShut {NoStop}%
\bibitem [{\citenamefont {Zanner}\ \emph {et~al.}(2021)\citenamefont {Zanner},
  \citenamefont {Orell}, \citenamefont {Schneider}, \citenamefont {Albert},
  \citenamefont {Oleschko}, \citenamefont {Juan}, \citenamefont {Silveri},\
  and\ \citenamefont {Kirchmair}}]{zanner2021coherent}%
  \BibitemOpen
  \bibfield  {author} {\bibinfo {author} {\bibfnamefont {M.}~\bibnamefont
  {Zanner}}, \bibinfo {author} {\bibfnamefont {T.}~\bibnamefont {Orell}},
  \bibinfo {author} {\bibfnamefont {C.~M.~F.}\ \bibnamefont {Schneider}},
  \bibinfo {author} {\bibfnamefont {R.}~\bibnamefont {Albert}}, \bibinfo
  {author} {\bibfnamefont {S.}~\bibnamefont {Oleschko}}, \bibinfo {author}
  {\bibfnamefont {M.~L.}\ \bibnamefont {Juan}}, \bibinfo {author}
  {\bibfnamefont {M.}~\bibnamefont {Silveri}}, \ and\ \bibinfo {author}
  {\bibfnamefont {G.}~\bibnamefont {Kirchmair}},\ }\href@noop {} {\enquote
  {\bibinfo {title} {Coherent control of a symmetry-engineered multi-qubit dark
  state in waveguide quantum electrodynamics},}\ } (\bibinfo {year} {2021}),\
  \Eprint {http://arxiv.org/abs/2106.05623} {arXiv:2106.05623 [quant-ph]}
  \BibitemShut {NoStop}%
\bibitem [{\citenamefont {Caneva}\ \emph {et~al.}(2015)\citenamefont {Caneva},
  \citenamefont {Manzoni}, \citenamefont {Shi}, \citenamefont {Douglas},
  \citenamefont {Cirac},\ and\ \citenamefont {Chang}}]{Caneva2015}%
  \BibitemOpen
  \bibfield  {author} {\bibinfo {author} {\bibfnamefont {T.}~\bibnamefont
  {Caneva}}, \bibinfo {author} {\bibfnamefont {M.~T.}\ \bibnamefont {Manzoni}},
  \bibinfo {author} {\bibfnamefont {T.}~\bibnamefont {Shi}}, \bibinfo {author}
  {\bibfnamefont {J.~S.}\ \bibnamefont {Douglas}}, \bibinfo {author}
  {\bibfnamefont {J.~I.}\ \bibnamefont {Cirac}}, \ and\ \bibinfo {author}
  {\bibfnamefont {D.~E.}\ \bibnamefont {Chang}},\ }\bibfield  {title} {\enquote
  {\bibinfo {title} {Quantum dynamics of propagating photons with strong
  interactions: a generalized input{\textendash}output formalism},}\ }\href
  {\doibase 10.1088/1367-2630/17/11/113001} {\bibfield  {journal} {\bibinfo
  {journal} {New J. Phys.}\ }\textbf {\bibinfo {volume} {17}},\ \bibinfo
  {pages} {113001} (\bibinfo {year} {2015})}\BibitemShut {NoStop}%
\bibitem [{\citenamefont {Schollw\"ock}(2011)}]{Schollwock2011}%
  \BibitemOpen
  \bibfield  {author} {\bibinfo {author} {\bibfnamefont {U.}~\bibnamefont
  {Schollw\"ock}},\ }\bibfield  {title} {\enquote {\bibinfo {title} {The
  density-matrix renormalization group in the age of matrix product states},}\
  }\href {\doibase https://doi.org/10.1016/j.aop.2010.09.012} {\bibfield
  {journal} {\bibinfo  {journal} {Annals of Physics}\ }\textbf {\bibinfo
  {volume} {326}},\ \bibinfo {pages} {96 -- 192} (\bibinfo {year}
  {2011})}\BibitemShut {NoStop}%
\bibitem [{\citenamefont {{Or{\'u}s}}(2014)}]{Orus2014}%
  \BibitemOpen
  \bibfield  {author} {\bibinfo {author} {\bibfnamefont {R.}~\bibnamefont
  {{Or{\'u}s}}},\ }\bibfield  {title} {\enquote {\bibinfo {title} {{A practical
  introduction to tensor networks: Matrix product states and projected
  entangled pair states}},}\ }\href {\doibase 10.1016/j.aop.2014.06.013}
  {\bibfield  {journal} {\bibinfo  {journal} {Annals of Physics}\ }\textbf
  {\bibinfo {volume} {349}},\ \bibinfo {pages} {117--158} (\bibinfo {year}
  {2014})}\BibitemShut {NoStop}%
\bibitem [{\citenamefont {Arranz~Regidor}\ \emph {et~al.}(2021)\citenamefont
  {Arranz~Regidor}, \citenamefont {Crowder}, \citenamefont {Carmichael},\ and\
  \citenamefont {Hughes}}]{Regidor2021}%
  \BibitemOpen
  \bibfield  {author} {\bibinfo {author} {\bibfnamefont {S.}~\bibnamefont
  {Arranz~Regidor}}, \bibinfo {author} {\bibfnamefont {G.}~\bibnamefont
  {Crowder}}, \bibinfo {author} {\bibfnamefont {H.}~\bibnamefont {Carmichael}},
  \ and\ \bibinfo {author} {\bibfnamefont {S.}~\bibnamefont {Hughes}},\
  }\bibfield  {title} {\enquote {\bibinfo {title} {Modeling quantum
  light-matter interactions in waveguide qed with retardation, nonlinear
  interactions, and a time-delayed feedback: Matrix product states versus a
  space-discretized waveguide model},}\ }\href {\doibase
  10.1103/PhysRevResearch.3.023030} {\bibfield  {journal} {\bibinfo  {journal}
  {Phys. Rev. Research}\ }\textbf {\bibinfo {volume} {3}},\ \bibinfo {pages}
  {023030} (\bibinfo {year} {2021})}\BibitemShut {NoStop}%
\bibitem [{\citenamefont {Lathauwer}\ \emph {et~al.}(2000)\citenamefont
  {Lathauwer}, \citenamefont {Moor},\ and\ \citenamefont
  {Vandewalle}}]{DeLathauwer2000}%
  \BibitemOpen
  \bibfield  {author} {\bibinfo {author} {\bibfnamefont {L.~D.}\ \bibnamefont
  {Lathauwer}}, \bibinfo {author} {\bibfnamefont {B.~D.}\ \bibnamefont {Moor}},
  \ and\ \bibinfo {author} {\bibfnamefont {J.}~\bibnamefont {Vandewalle}},\
  }\bibfield  {title} {\enquote {\bibinfo {title} {A multilinear singular value
  decomposition},}\ }\href {\doibase 10.1137/s0895479896305696} {\bibfield
  {journal} {\bibinfo  {journal} {{SIAM} Journal on Matrix Analysis and
  Applications}\ }\textbf {\bibinfo {volume} {21}},\ \bibinfo {pages}
  {1253--1278} (\bibinfo {year} {2000})}\BibitemShut {NoStop}%
\bibitem [{\citenamefont {Vervliet}\ \emph {et~al.}(2016)\citenamefont
  {Vervliet}, \citenamefont {Debals}, \citenamefont {Sorber}, \citenamefont
  {Van~Barel},\ and\ \citenamefont {De~Lathauwer}}]{tensorlab}%
  \BibitemOpen
  \bibfield  {author} {\bibinfo {author} {\bibfnamefont {N.}~\bibnamefont
  {Vervliet}}, \bibinfo {author} {\bibfnamefont {O.}~\bibnamefont {Debals}},
  \bibinfo {author} {\bibfnamefont {L.}~\bibnamefont {Sorber}}, \bibinfo
  {author} {\bibfnamefont {M.}~\bibnamefont {Van~Barel}}, \ and\ \bibinfo
  {author} {\bibfnamefont {L.}~\bibnamefont {De~Lathauwer}},\ }\href
  {https://www.tensorlab.net} {\enquote {\bibinfo {title} {Tensorlab 3.0},}\ }
  (\bibinfo {year} {2016}),\ \bibinfo {note} {available online.}\BibitemShut
  {Stop}%
\bibitem [{\citenamefont {Eisert}\ \emph {et~al.}(2010)\citenamefont {Eisert},
  \citenamefont {Cramer},\ and\ \citenamefont {Plenio}}]{Eisert2010}%
  \BibitemOpen
  \bibfield  {author} {\bibinfo {author} {\bibfnamefont {J.}~\bibnamefont
  {Eisert}}, \bibinfo {author} {\bibfnamefont {M.}~\bibnamefont {Cramer}}, \
  and\ \bibinfo {author} {\bibfnamefont {M.~B.}\ \bibnamefont {Plenio}},\
  }\bibfield  {title} {\enquote {\bibinfo {title} {\textit{Colloquium:} area
  laws for the entanglement entropy},}\ }\href {\doibase
  10.1103/RevModPhys.82.277} {\bibfield  {journal} {\bibinfo  {journal} {Rev.
  Mod. Phys.}\ }\textbf {\bibinfo {volume} {82}},\ \bibinfo {pages} {277--306}
  (\bibinfo {year} {2010})}\BibitemShut {NoStop}%
\bibitem [{\citenamefont {Zhang}\ \emph {et~al.}(2015)\citenamefont {Zhang},
  \citenamefont {Xie}, \citenamefont {Wang}, \citenamefont {Poddubny},
  \citenamefont {Lu}, \citenamefont {Wang}, \citenamefont {Gu}, \citenamefont
  {Liu}, \citenamefont {Xu}, \citenamefont {Shen}, \citenamefont {Rubo},
  \citenamefont {Altshuler}, \citenamefont {Kavokin},\ and\ \citenamefont
  {Chen}}]{Zhang2015}%
  \BibitemOpen
  \bibfield  {author} {\bibinfo {author} {\bibfnamefont {L.}~\bibnamefont
  {Zhang}}, \bibinfo {author} {\bibfnamefont {W.}~\bibnamefont {Xie}}, \bibinfo
  {author} {\bibfnamefont {J.}~\bibnamefont {Wang}}, \bibinfo {author}
  {\bibfnamefont {A.}~\bibnamefont {Poddubny}}, \bibinfo {author}
  {\bibfnamefont {J.}~\bibnamefont {Lu}}, \bibinfo {author} {\bibfnamefont
  {Y.}~\bibnamefont {Wang}}, \bibinfo {author} {\bibfnamefont {J.}~\bibnamefont
  {Gu}}, \bibinfo {author} {\bibfnamefont {W.}~\bibnamefont {Liu}}, \bibinfo
  {author} {\bibfnamefont {D.}~\bibnamefont {Xu}}, \bibinfo {author}
  {\bibfnamefont {X.}~\bibnamefont {Shen}}, \bibinfo {author} {\bibfnamefont
  {Y.~G.}\ \bibnamefont {Rubo}}, \bibinfo {author} {\bibfnamefont {B.~L.}\
  \bibnamefont {Altshuler}}, \bibinfo {author} {\bibfnamefont {A.~V.}\
  \bibnamefont {Kavokin}}, \ and\ \bibinfo {author} {\bibfnamefont
  {Z.}~\bibnamefont {Chen}},\ }\bibfield  {title} {\enquote {\bibinfo {title}
  {Weak lasing in one-dimensional polariton superlattices},}\ }\href {\doibase
  10.1073/pnas.1502666112} {\bibfield  {journal} {\bibinfo  {journal}
  {Proceedings of the National Academy of Sciences}\ }\textbf {\bibinfo
  {volume} {112}},\ \bibinfo {pages} {E1516} (\bibinfo {year}
  {2015})}\BibitemShut {NoStop}%
\bibitem [{\citenamefont {Nalitov}\ \emph {et~al.}(2017)\citenamefont
  {Nalitov}, \citenamefont {Liew}, \citenamefont {Kavokin}, \citenamefont
  {Altshuler},\ and\ \citenamefont {Rubo}}]{Nalitov2017}%
  \BibitemOpen
  \bibfield  {author} {\bibinfo {author} {\bibfnamefont {A.~V.}\ \bibnamefont
  {Nalitov}}, \bibinfo {author} {\bibfnamefont {T.~C.~H.}\ \bibnamefont
  {Liew}}, \bibinfo {author} {\bibfnamefont {A.~V.}\ \bibnamefont {Kavokin}},
  \bibinfo {author} {\bibfnamefont {B.~L.}\ \bibnamefont {Altshuler}}, \ and\
  \bibinfo {author} {\bibfnamefont {Y.~G.}\ \bibnamefont {Rubo}},\ }\bibfield
  {title} {\enquote {\bibinfo {title} {Spontaneous polariton currents in
  periodic lateral chains},}\ }\href {\doibase 10.1103/PhysRevLett.119.067406}
  {\bibfield  {journal} {\bibinfo  {journal} {Phys. Rev. Lett.}\ }\textbf
  {\bibinfo {volume} {119}},\ \bibinfo {pages} {067406} (\bibinfo {year}
  {2017})}\BibitemShut {NoStop}%
\bibitem [{\citenamefont {Brehm}\ \emph {et~al.}(2021)\citenamefont {Brehm},
  \citenamefont {Poddubny}, \citenamefont {Stehli}, \citenamefont {Wolz},
  \citenamefont {Rotzinger},\ and\ \citenamefont
  {Ustinov}}]{brehm2020waveguide}%
  \BibitemOpen
  \bibfield  {author} {\bibinfo {author} {\bibfnamefont {J.~D.}\ \bibnamefont
  {Brehm}}, \bibinfo {author} {\bibfnamefont {A.~N.}\ \bibnamefont {Poddubny}},
  \bibinfo {author} {\bibfnamefont {A.}~\bibnamefont {Stehli}}, \bibinfo
  {author} {\bibfnamefont {T.}~\bibnamefont {Wolz}}, \bibinfo {author}
  {\bibfnamefont {H.}~\bibnamefont {Rotzinger}}, \ and\ \bibinfo {author}
  {\bibfnamefont {A.~V.}\ \bibnamefont {Ustinov}},\ }\bibfield  {title}
  {\enquote {\bibinfo {title} {Waveguide bandgap engineering with an array of
  superconducting qubits},}\ }\href {\doibase 10.1038/s41535-021-00310-z}
  {\bibfield  {journal} {\bibinfo  {journal} {npj Quantum Materials}\ }\textbf
  {\bibinfo {volume} {6}},\ \bibinfo {pages} {10} (\bibinfo {year}
  {2021})}\BibitemShut {NoStop}%
\bibitem [{\citenamefont {Lalumi\`ere}\ \emph {et~al.}(2013)\citenamefont
  {Lalumi\`ere}, \citenamefont {Sanders}, \citenamefont {van Loo},
  \citenamefont {Fedorov}, \citenamefont {Wallraff},\ and\ \citenamefont
  {Blais}}]{Blais2013}%
  \BibitemOpen
  \bibfield  {author} {\bibinfo {author} {\bibfnamefont {K.}~\bibnamefont
  {Lalumi\`ere}}, \bibinfo {author} {\bibfnamefont {B.~C.}\ \bibnamefont
  {Sanders}}, \bibinfo {author} {\bibfnamefont {A.~F.}\ \bibnamefont {van
  Loo}}, \bibinfo {author} {\bibfnamefont {A.}~\bibnamefont {Fedorov}},
  \bibinfo {author} {\bibfnamefont {A.}~\bibnamefont {Wallraff}}, \ and\
  \bibinfo {author} {\bibfnamefont {A.}~\bibnamefont {Blais}},\ }\bibfield
  {title} {\enquote {\bibinfo {title} {Input-output theory for waveguide qed
  with an ensemble of inhomogeneous atoms},}\ }\href {\doibase
  10.1103/PhysRevA.88.043806} {\bibfield  {journal} {\bibinfo  {journal} {Phys.
  Rev. A}\ }\textbf {\bibinfo {volume} {88}},\ \bibinfo {pages} {043806}
  (\bibinfo {year} {2013})}\BibitemShut {NoStop}%
\bibitem [{\citenamefont {Ye}\ \emph {et~al.}(2019)\citenamefont {Ye},
  \citenamefont {Ge}, \citenamefont {Wu}, \citenamefont {Wang}, \citenamefont
  {Gong}, \citenamefont {Zhang}, \citenamefont {Zhu}, \citenamefont {Yang},
  \citenamefont {Li}, \citenamefont {Liang}, \citenamefont {Lin}, \citenamefont
  {Xu}, \citenamefont {Guo}, \citenamefont {Sun}, \citenamefont {Cheng},
  \citenamefont {Ma}, \citenamefont {Meng}, \citenamefont {Deng}, \citenamefont
  {Rong}, \citenamefont {Lu}, \citenamefont {Peng}, \citenamefont {Fan},
  \citenamefont {Zhu},\ and\ \citenamefont {Pan}}]{Ye2019}%
  \BibitemOpen
  \bibfield  {author} {\bibinfo {author} {\bibfnamefont {Y.}~\bibnamefont
  {Ye}}, \bibinfo {author} {\bibfnamefont {Z.-Y.}\ \bibnamefont {Ge}}, \bibinfo
  {author} {\bibfnamefont {Y.}~\bibnamefont {Wu}}, \bibinfo {author}
  {\bibfnamefont {S.}~\bibnamefont {Wang}}, \bibinfo {author} {\bibfnamefont
  {M.}~\bibnamefont {Gong}}, \bibinfo {author} {\bibfnamefont {Y.-R.}\
  \bibnamefont {Zhang}}, \bibinfo {author} {\bibfnamefont {Q.}~\bibnamefont
  {Zhu}}, \bibinfo {author} {\bibfnamefont {R.}~\bibnamefont {Yang}}, \bibinfo
  {author} {\bibfnamefont {S.}~\bibnamefont {Li}}, \bibinfo {author}
  {\bibfnamefont {F.}~\bibnamefont {Liang}}, \bibinfo {author} {\bibfnamefont
  {J.}~\bibnamefont {Lin}}, \bibinfo {author} {\bibfnamefont {Y.}~\bibnamefont
  {Xu}}, \bibinfo {author} {\bibfnamefont {C.}~\bibnamefont {Guo}}, \bibinfo
  {author} {\bibfnamefont {L.}~\bibnamefont {Sun}}, \bibinfo {author}
  {\bibfnamefont {C.}~\bibnamefont {Cheng}}, \bibinfo {author} {\bibfnamefont
  {N.}~\bibnamefont {Ma}}, \bibinfo {author} {\bibfnamefont {Z.~Y.}\
  \bibnamefont {Meng}}, \bibinfo {author} {\bibfnamefont {H.}~\bibnamefont
  {Deng}}, \bibinfo {author} {\bibfnamefont {H.}~\bibnamefont {Rong}}, \bibinfo
  {author} {\bibfnamefont {C.-Y.}\ \bibnamefont {Lu}}, \bibinfo {author}
  {\bibfnamefont {C.-Z.}\ \bibnamefont {Peng}}, \bibinfo {author}
  {\bibfnamefont {H.}~\bibnamefont {Fan}}, \bibinfo {author} {\bibfnamefont
  {X.}~\bibnamefont {Zhu}}, \ and\ \bibinfo {author} {\bibfnamefont {J.-W.}\
  \bibnamefont {Pan}},\ }\bibfield  {title} {\enquote {\bibinfo {title}
  {Propagation and localization of collective excitations on a 24-qubit
  superconducting processor},}\ }\href {\doibase
  10.1103/PhysRevLett.123.050502} {\bibfield  {journal} {\bibinfo  {journal}
  {Phys. Rev. Lett.}\ }\textbf {\bibinfo {volume} {123}},\ \bibinfo {pages}
  {050502} (\bibinfo {year} {2019})}\BibitemShut {NoStop}%
\bibitem [{\citenamefont {Zhang}\ \emph {et~al.}(2020)\citenamefont {Zhang},
  \citenamefont {Yu},\ and\ \citenamefont {M\o{}lmer}}]{Zhang2019arXiv}%
  \BibitemOpen
  \bibfield  {author} {\bibinfo {author} {\bibfnamefont {Y.-X.}\ \bibnamefont
  {Zhang}}, \bibinfo {author} {\bibfnamefont {C.}~\bibnamefont {Yu}}, \ and\
  \bibinfo {author} {\bibfnamefont {K.}~\bibnamefont {M\o{}lmer}},\ }\bibfield
  {title} {\enquote {\bibinfo {title} {Subradiant bound dimer excited states of
  emitter chains coupled to a one dimensional waveguide},}\ }\href {\doibase
  10.1103/PhysRevResearch.2.013173} {\bibfield  {journal} {\bibinfo  {journal}
  {Phys. Rev. Research}\ }\textbf {\bibinfo {volume} {2}},\ \bibinfo {pages}
  {013173} (\bibinfo {year} {2020})}\BibitemShut {NoStop}%
\bibitem [{\citenamefont {Poddubny}(2020)}]{Poddubny2019quasiflat}%
  \BibitemOpen
  \bibfield  {author} {\bibinfo {author} {\bibfnamefont {A.~N.}\ \bibnamefont
  {Poddubny}},\ }\bibfield  {title} {\enquote {\bibinfo {title} {Quasiflat band
  enabling subradiant two-photon bound states},}\ }\href {\doibase
  10.1103/PhysRevA.101.043845} {\bibfield  {journal} {\bibinfo  {journal}
  {Phys. Rev. A}\ }\textbf {\bibinfo {volume} {101}},\ \bibinfo {pages}
  {043845} (\bibinfo {year} {2020})}\BibitemShut {NoStop}%
\bibitem [{\citenamefont {Zhang}\ and\ \citenamefont
  {M\o{}lmer}(2020)}]{Zhang2020d}%
  \BibitemOpen
  \bibfield  {author} {\bibinfo {author} {\bibfnamefont {Y.-X.}\ \bibnamefont
  {Zhang}}\ and\ \bibinfo {author} {\bibfnamefont {K.}~\bibnamefont
  {M\o{}lmer}},\ }\bibfield  {title} {\enquote {\bibinfo {title} {Subradiant
  emission from regular atomic arrays: Universal scaling of decay rates from
  the generalized {B}loch theorem},}\ }\href {\doibase
  10.1103/PhysRevLett.125.253601} {\bibfield  {journal} {\bibinfo  {journal}
  {Phys. Rev. Lett.}\ }\textbf {\bibinfo {volume} {125}},\ \bibinfo {pages}
  {253601} (\bibinfo {year} {2020})}\BibitemShut {NoStop}%
\end{thebibliography}

%

\end{document}